\documentclass[manuscript]{aastex}

\usepackage{txfonts}
\usepackage{lscape}
\shorttitle{Saturation in Chromospheric Emission}
\shortauthors{Christian et al.}

\begin{document}

\title{The Search for Super-saturation in Chromospheric Emission}
\vskip1.0truecm

\author{Damian J. Christian}
\affil{Department of Physics and Astronomy, California State University, \\
18111 Nordhoff Street, Northridge, CA
91330; daman.christian@csun.edu} 
\author{Mihalis Mathioudakis} 
\affil{Astrophysics Research Centre, School of Mathematics and Physics, Queen's University, Belfast, BT7~1NN, Northern Ireland, UK}
\author{Tersi Arias}
\affil{Department of Physics and Astronomy, California State University, \\
18111 Nordhoff Street, Northridge, CA
91330}

\author{Moira Jardine}
\affil{School of Physics and Astronomy, North Haugh,  St. Andrews KY169SS, Scoctland, UK}

\author{David Jess}
\affil{Astrophysics Research Centre, School of Mathematics and Physics, Queen's University, Belfast, BT7~1NN, Northern Ireland, UK}


\begin{abstract}

We investigate if the super-saturation phenomenon observed at X-ray wavelengths for the corona, exists in the chromosphere for rapidly rotating late-type stars.
Moderate resolution optical spectra of fast rotating EUV- and X-ray- selected 
late-type stars were obtained.  Stars in $\alpha$ Per were observed in the northern hemisphere with the Isaac Newton 2.5 m telescope and IDS spectrograph. Selected objects from IC 2391 and IC 2602 were observe in the southern hemisphere with the Blanco 4m telescope and R-C spectrograph at CTIO.
Ca II H \& K fluxes were measured for all stars in our sample. We find the saturation level for Ca II K at  log($L_{CaK}/L_{bol}$ ) = -4.08. The Ca II K flux does not show a decrease as a function of increased rotational velocity or smaller Rossby number as observed in the X-ray.
This lack of "super-saturation" supports the idea of coronal-stripping as the cause of saturation and super-saturation in stellar chromospheres and corona, but the detailed underlying mechanism is still under investigation.

\end{abstract}


\section{Introduction}
The Sun's chromosphere is a complex region where pressure dominance at the lower photosphere becomes magnetic dominance at high altitudes in the atmosphere. Energy dissipation in the chromosphere and the associated heating processes have been a topic of intense debate.
Virtually all late-type stars with surface convective envelopes (types F to M) have chromospheres and it has been well established that the strength of the dynamo action
increases with stellar rotation rates \citep{S72}. A better understanding of the 
Sun's outer atmosphere can help us to understand other stars while stars of different age and mass can provide a better understanding of the evolution of the solar dynamo. 

In recent years, much observational effort has concentrated
on the study of open clusters \citep{S98, J99}. In the younger
clusters ($\alpha$ Per, 50Myr) a large number of G and K dwarfs are
rotating rapidly ($\upsilon$ sin $i$ $\geq$ 100 km s$^{-1}$), whereas in the
older Hyades and in the solar neighborhood rapid rotation is confined
to the M dwarfs.
When comparing the activity properties of stars covering a large range of masses, 
the Rossby number N$_R$, (ratio of rotation period to convective
turnover time $\tau_c$) has often been used instead of the rotational period,
\citep{SF87, M95}.
This is because, in addition to the dynamo's dependence rotation, there is an increase in the convective turnover time 
for lower mass stars \citep{KD96}.

The Rossby diagram for open clusters has shown that the "plateau" of coronal
saturation occurs for L$_X$/L$_{bol} \approx 10^{-3} - 10^{-4}$ \citep{S97, Q98, P03}. 
Saturation can be understood in terms of the limited area of the stellar surface. 
As more surface field is generated with increased rotation and
deep convection, a point is eventually reached where the stellar surface is totally covered with active regions. The foot-points of magnetic loops crowd each other out and no more new loops can be generated. Consequently, the heating reaches a maximum and so does the radiative emission. In that scenario, the emission would be expected to scale with surface area. A second explanation for the saturation is based on the feedback between the induced fields and velocities in the convection zone. As an increased angular velocity generates stronger magnetic fields, a point is eventually reached where the ratio between the total magnetic energy and kinetic energy approaches unity. 
The equipartition achieved provides an upper boundary to the total magnetic energy, since the Lorentz forces ({\bf jxB}, the vector cross product of the current {\bf j}, and magnetic field, {\bf B})  suppress convection.
The strong Lorentz forces help in reducing differential rotation and therefore its ability to induce toroidal field from the poloidal. As a result limited atmospheric emission will be created and is expected to be a function of parameters such as the stellar radius 
\citep{G85}.    
In general chromospheric saturation has been found to occur at similar Rossby numbers as coronal saturation \citep{M09}  and saturation of the coronal emission has been interpreted as saturation of the dynamo itself \citep{VW87} . However, \citet{M95} found some evidence for coronal saturation to occur at
lower Rossby numbers (higher rotation rates) than chromospheric saturation.
%
\citet{R09}
find saturation of the magnetic flux  can occur at lower rotational velocities for M-dwarfs and understanding saturation in these different regimes remains a challenge to both observations and theories.

Several extremely fast rotators have been observed to show a decline in their coronal emission with increasing rotation. This effect has been termed
{\em super-saturation} 
\citep{P96, S97, Q98, J00}.  One explanation for this effect has been given as resulting from the reductiion of the X-ray emitting volume of 
rapid rotators  by centrifugal stripping \citep{J04}.
The increased rotational velocity will decrease the apparent surface gravity, and cause  
increased pressure in the outer parts of large coronal loops. The magnetic field lines get distorted and eventually break open leading to saturation. Combined with a saturated dynamo these processes will ultimately lead to a reduction in the coronal X-ray emission. Recently, \citet{J11} have recently shown support for the centrifugal stripping idea for explaining  super-saturation using XMM-Newton observations of K and M stars in NGC~2547. Thus for super-saturated stars, 
the hot coronal plasma will become unstable and cool down to chromospheric temperatures  \citep{A86, S99}.
More cool loops emitting in chromospheric and transition region lines 
can exist but not in X-rays. This would imply that there should be no evidence for super-saturation at chromospheric temperatures. 



Recent work has supported this expected lack of supersaturation in the chromosphere. Observing the Ca IR triplet,  \citet{M09} found no evidence supersturation in the chromosphere for young open clusters IC~2391 and IC~2602.  Such studies were extended to rapidly rotating M dwarfs in NGC~2516 \citep{JJ10}, which also found no decrease in chromospheric emission even at very low Rossby numbers ($\lesssim$ 0.01). 

The present paper uses moderate resolution spectra for a sample of fast rotating late-type stars  to search for the effects of super-saturation in their chromospheres. A description of the sample selection criteria is given in Section 2.1. 
In $\S$~2.2 we describe the observations, instrumental set-up and data reduction techniques.
In $\S$~3 we present our results including measurements of the chromospheric
emission in terms of Ca II K and make a comparison with the 
rotational velocities and Rossby numbers, and in $\S$ 4, 
we discuss trends for these parameters and compare the results to super-saturation models for the chromosphere. 
Lastly, in $\S$ 5 we summarize our findings. 

\section{Observations \& Data Reduction}
\subsection{Sample Selection}
\noindent
Our samples consists of 72 active late-type stars with projected equatorial rotation velocities in the range of 20 - 200 km sec$^{-1}$.  Our northern sample concentrated on stars in $\alpha$ Per \citep{R96} 
while our  southern sample includes stars from IC 2391  \citep{PS96, S97}
and IC 2602  \citep{R95, S97}.
The optical identification campaigns that followed X-ray and EUV surveys provided some additional {\it solar neighborhood} targets  \citep{J98, CC01, CM02, Hu04}.
 V magnitudes versus the color index, $B-V$ are shown for our sample in Figure 1a, and $V-K$ computed from K colors from the 2MASS catalog is shown in Fig 1b. The reduced scatter demonstrates that V-K is a more suitable color for these late spectral types.

\subsection{Optical Observations \& Analysis}

The southern objects were observed using the CTIO Blanco 4-meter telescope and R-C spectrograph, while in the north we used the 2.5-meter Isaac Newton Telescope equipped with the Intermediate Dispersion Spectrograph (IDS).   The CTIO observations were obtained on 2 nights in 2005 Feb 23-24. The KPGL1 grating was used giving a resolution of 1\AA\ per pixel and covering the  range of $\approx$3300 - 5800\AA.  Spectra were bias subtracted and flat fielded with an internal quartz lamp, and the wavelength calibration was established with HeNeAr lamps.
The  Isaac Newton Group of Telescopes (INT) spectra were obtained in 2007 December 23-27 using the R1200B grating. Our set-up covered the 3600 to 4600 \AA\ range with a spectral resolution of 0.47~\AA/pixel.  Wavelength calibration was established using CuAr plus CuNe lamps before and after each exposure. Flat fields were obtained with a quartz lamp.
Several radial velocity standards and flux standards were 
 used to establish the accuracy of the velocity scale and flux calibration respectively. 
The observations were reduced using standard routines within IRAF.   Two dimensional images were bias subtracted and flat fielded with the {\em ccdred} package and wavelength and flux calibrated with the {\it refspec} and {\it calibrate} routines within noao.onedspec, respectively. 

The Ca II H \& K line fluxes were derived by fitting Gaussians to the flux calibrated spectra as described below. We focus on the Ca II K line at $\lambda\lambda$3933 due to the possible blending of Ca II H with H$\epsilon$.  
We show a sample of sources with both strong Ca II H \& K in emission in Figure 2.

To estimate the photospheric contribution to the Ca II line fluxes we constructed stellar models using the {\em Spectroscopy Made Easy} (SME) code \citep{VP96}. 
Models were created from 3500 to 6500 K in 300 K steps for solar metallicities and
 gravity log g = 4.5. Additional models were created at each temperature for appropriate values of $vsini$ from 0 to 200 km s$^{-1}$.  
 The normalization and systemic radial velocity $v_{rad}$ were left as free parameters. 
Atomic line data were obtained from the Vienna Atomic Line Database (VALD) \citep{PK95}. 
The SME models were then subtracted from each observed spectrum using the model closest in temperature (T) and $v sini$. 
The difference between the observed (chromospheric contribution) and SME model (photospheric contribution) were then re-fit with a new Gaussian. 
Spectra for representative stars at four different temperature are shown in Figure 3.  In general we find the photosphere contribution is about $\approx$ 50\%.  However, 
photospheric contributions for the latest spectral types (T $<$ 4000) are in general smaller and on the order of 10\%. 
Typical photospheric contribution in the temperature range between 4000 and 5500 K are  $\approx$ 50\%, but can be as high as a factor of $\approx$2--3 for stars with temperatures above 5500K and for those with the weakest emission. 
In general the photospheric contribution will depend not only on a star's temperature, but its chromospheric activity level and rate of rotation. 
We note, any photospheric contribution not accounted for will only act to increase the Ca fluxes, not decrease them. 

Surface fluxes are calculated using the relationship between observed flux and surface flux given in  \citep{O82, Ru89, MD89, MD92}. Reddening values  for $\alpha$ Per were taken
from \citet{R96} with  E($B-V$) = 0.1, and from \citet{PS96} for IC~2391 and \citet{R95} for IC~2602 with  E($B-V$) = 0.006 and 0.04, respectively.
Bolometric corrections were taken from \citet{J66} and \citet{B91}. 
Rotational velocities ($ v$ sin$i$) were taken from the literature, and rotational periods, P, were derived from these using $P = 2\pi R_*/v sini$, where R$_*$ is the stellar radius from the ratio of the stellar flux to observed flux as given in equation 1 in \citet{MD92}. Possible sources of error in the periods from the stellar radii are treated in \S\ 3.
Marsden et al. 2009 have shown rotational velocity values for IC 2391 and IC 2602 from the literature to be in good agreement with their derived values. 
Previously \citep{CM02} we had used relations for Rossby number and $B-V$ from \citep{N84}. However these relations are flat for later type stars with $B-V >$ 1 (see discussion in \citet{J11}), and we calculated the Rossby number using the 
relation:

\begin{equation}
log(N_R) = log(P) - 1.1 + 0.5 log(L_{bol}/L_{\odot})
\end{equation}

from \citet{J11}. Where N$_R$ is the Rossby number,  P is the stellar rotation period, L$_{bol}$ is the bolometric luminosity and  L$_{\odot}$ is the solar bolometric luminosity. The scale factor 1.1 is set for the log of the convective turnover time for solar type stars \citep{N84,J11}.

\section{Results}

Approximately 85\% of the stars in our sample showed Ca II H \& K in emission. The remaining 15\% had no detectable Ca II H \& K in emission. These objects are of the earlier spectral types, $B-V$ $<$ 0.6, and the non-detection is most likely due to the combination of strong photospheric continuum and spectral  resolution rather than the absence of a chromosphere.  The Ca II H \&K line fluxes and stellar parameters are shown in Table~1\label{tab1}. 
We computed the $L_{CaK}/L_{bol}$ for the stars in our sample using the relationship between observed flux and surface flux  \citep{Ru89,MD89}.

The X-ray emission of coronally active late-type stars ÒsaturatesÓ at  $\approx$ 1 part in 1000 of the starÕs bolometric luminosity ($L_X/L_{bol}$). Many stars in our sample have $L_X/L_{bol}$ between 10$^{-3}$ and 10$^{-4}$.  
We compare the X-ray and Ca II K luminosity ratios in Figure 4. 
In the figure we have  
indicated stars showing super-saturation in X-rays from \citet{J00} ($L_X/L_{bol}$ and vsin$i$ $>$ 90 
 km s$^{-1}$), and the figure shows a slight trend that stars with higher chromospheric luminosity have lower X-ray luminosities. The $L_{Ca K}$ luminosities for the supersaturated stars differ from the entire sample with a modest, but significant Kolmogorov-Smirnov statistic of 0.6. 

In Figure 5 we plot the $L_{Ca K}/L_{bol}$ as a function of v$sini$. 
The average log($L_{Ca K}/L_{bol}$)  is $-$4.08. We have estimated the errors for log($L_{Ca K}/L_{bol}$) combining a 10\% uncertainty in the flux estimates and an additional 10\% possible systematic error in converting from counts to flux along with any uncertainty in radii used in calculating the bolometric luminosities (generally less than 10\%).
Combining these errors, we conservatively estimate a 30\% uncertainty in $L_{Ca K}/L_{bol}$ and  show a sample error of 30\% in Figure 5. Although there is significant scatter below 50 km~s$^{-1}$, there is no evidence for a decrease in $L_{Cak}/L_{bol}$ for the highest rotational velocities and this is treated in \S\ 4. The low point in the figure, at 
$\approx$170~km~s$^{-1}$ is AP~139, which had an anomalously low Ca II K flux for its rotational velocity, but its rotational velocity was determined from its photometric period \citep{S97}, and may actually be a binary.

To compare the rotation and activity of our sample of stars with different masses and radii it is convenient to use the Rossby number 
N$_R$, (ratio of rotation period to convective 
turnover time $\tau_c$) \citep{J11}.  Periods were derived from rotational velocities and stellar radius calculated from \citet{MD92} as described in Section 2.2. A 20\% error in the stellar radii  will cause a 20\% error in derived periods,
but this only translates to a change in log(N$_R$) of 0.08.
The saturation levels of coronal X-rays are reached at log(N$_R$) $\approx -1$ ($L_X/L_{bol}$ $\approx$ 10$^{-3}$) and extend to log(N$_R$) $\approx -1.8$  \citep{J11}. Similar ranges were found for pre-main sequence stars by \citet{F03}.
 The discovery  of Òsuper-saturationÓ at the highest stellar rotation rates 
\citep{P96, R96, S97}, 
was recently confirmed using a sample of 
G  \&  K dwarfs by \citet{J11}. The same authors do not find strong evidence for super-saturation in M dwarfs.  
Some evidence for super-saturation has also been found in the extreme-ultraviolet  \citep{CM02}. 

Our sample has 26 stars with the log of the Rossby number, log(N$_R$) less than $-$1.6. In Figure 6 we plot our Ca II K luminosities  ($L_{CaK}/L_{bol}$)
as a function of Rossby number. Although there is a large amount of scatter in $L_{CaK}/L_{bol}$ for the lowest Rossby numbers, there is no obvious decrease in $L_{CaK}/L_{bol}$ for the fastest rotators (log(N$_R$)  $< - 1.8$).  We over-plot (dotted line) linear fits to the X-ray observations from \citet{J11} in the 3 regimes: log(N$_R$)  $<  -1.8$, $-1.8 <$ log(N$_R$) $< -0.8$, and log(N$_R$) $> -$0.8, where we used the slopes in these regions and set the saturated region to the mean of $L_{CaK}/L_{bol}$. 
We note, the over-plotted loci are not formally correct for the latest-type stars in our sample, but are included for comparison.
We have converted the published rotational velocities to periods for calculations of the Rossby numbers and this may add some additional uncertainty, but in the x-direction, not in $L_{Ca K}/L_{bol}$. We show a sample error of log(N$_R$) $\approx$ 0.08 in Figure 6. For example, using the relation for Rossby from \citet{N84} generally finds values 0.2 higher than the \cite{J11} relation shown in Figure 6, but the \citet{N84} relation is not correct for stars with $B-V$ $>$ 1, as discussed in Section 2.2.

\section{Discussion: $L_{CaK}/L_{bol}$ and Chromospheric Emission}
\citet{M09} found saturation to set in near log(Rossby) = $-$0.8 using observations of the Ca II  IR triplet  for the fast rotators in IC 2391 and IC 2602. Although our Rossby diagram  (Fig~6) has much scatter, we find a saturation level for the Ca K emission at 
log($L_{CaK}/L_{bol}$) of $-$4.08 and to be consistent with setting in at  
log($N_R$) of $\approx$ $-$0.8, although given the limited sample and scatter, the range could be between log($N_R$) of $\approx$ $-$0.6 to $-$1.0.

The coronal emission of very fast rotating stars begins to decrease at a v sini $>$ 100 km~s$^{-1}$, Rossby number of $\approx$ $-$1.6. 
Several authors,  \citep{S97, Q98, J00}
have observed that coronal emission decreased for the fastest rotating stars (v sini $>$ 100 km~s$^{-1}$) and at a Rossby number of $\approx$ $-$1.6. 
No significant decrease of $L_{CaK}/L_{bol}$ with higher rotational velocities or lower 
Rossby numbers was observed in our sample.  The mean $L_{CaK}/L_{bol}$ was 8.76 $\times$ 10$^{-5}$ for stars with $vsini$ less than 100 km~s$^{-1}$ and the average for stars with vsini greater than 100 km~s$^{-1}$ was 9.68 $\times$ 10$^{-5}$.  These two values differ by $\approx$10\% and are well within uncertainties. %
For comparison, the X-ray sample of K \& M stars show  decrease of the mean log($L_X/L_{bol}$) from $-$3.0  for the complete sample to $-$3.5 for stars at log(Rossby)            $\lesssim$ $-$2 \citep{J11}.
\citet{M09} find a similar decrease for the X-ray saturation levels 
 for the fastest rotators.  However, here we have found no evidence for the decrease in the chromospheric emission of Ca II K.

 \citet{M09}
suggest that saturation in the chromosphere (8542\AA) and corona occur at a Rossby number of $\approx$ $-$1. 
The Mg II results from \citet{CC07} suggest that saturation occurs at 
log(N$_R$) of $\approx$ $-$0.7. However,  \citet{M09} argue these are possibly still within the error bars. %
Also, the results of  \citet{F03} and \citet{J00} find  that the 
saturation levels for 
the corona may occur as early as a Rossby value of $-$0.7. Although we agree these results may be within 
the errors for saturation at Rossby values of $-$1 seen for the corona, we can then ask: {\it What does this mean if the corona actually reaches saturation at a higher Rossby number (slower rotation rate) than the chromosphere?} This may be related to the fact that the co-rotation radius is larger for slower rotators and the onset of saturation of the dynamo can occur at higher Rossby number in both the corona and chromosphere, although the exact mechanism for saturation in presently not well understood.

An explanation for the saturation of coronal X-rays has been given as
the reduction of the X-ray emitting volume of late-type stars due to centrifugal stripping 
\citep{J00, JU99, M09}. 
The increased rotational velocity decreases the apparent surface 
gravity and the centrifugal forces increase the pressure and density in the loop 
summits. This outward pressure may eventually force
the largest closed loops to open up, releasing their plasma into the 
stellar wind. This process would reduce the volume of confined plasma in the corona 
and would therefore reduce the overall X-ray emission, potentially masking 
the onset of dynamo saturation.  Support for the coronal stripping model has been found by \citet{J04}. 

An additional effect of rapid rotation has also been suggested
by \citet{A86} and \citet{S99}.
Since at any given height the plasma pressure decreases with decreasing
temperature, low temperature loops have a lower pressure than high 
temperature loops and are therefore less susceptible to centrifugal stripping. If rapid 
rotators were to possess a higher proportion of cool
loops, these could remain closed out to heights where hotter loops 
would have been opened up by their internal pressure. These cool loops might contribute to the UV and EUV radiation, but not to X-rays. The shape of the emission 
measure distribution with temperature would therefore shift, with more emission appearing at  lower temperatures.
Thus, if the coronal heating mechanism in
rapid rotators were to favor the production of a greater fraction of 
cool loops than in slowly-rotating stars, this might explain the saturation of X-ray 
emission measure with increasing rotation rate.
%
%
This would also imply that there should be no evidence for super-saturation at chromospheric temperatures and this agrees well with our results found here for Ca II H \& K and recent results for observations of the Ca IR triplet \citep{M09, JJ10}. 

\section{Conclusions}
Some extremely fast rotating stars show a decline in their coronal emission
with increasing rotation. This effect has been termed
{\em super-saturation} and as coronal X-rays cool down to chromospheric temperatures more cool loops emitting in chromospheric and transition region lines 
can exist but not in X-rays. We have investigated this idea by observing the Ca II H\& K emission from a sample of rapidly rotating stars in both the northern and southern hemispheres.  Our moderate resolution 
optical observations find a  chromospheric saturation level of $L_{CaK}/L_{bol}$  = -4.08. Comparison of $L_{CaK}/L_{bol}$  with $v$sin$i$ shows no significant decrease for the fastest rotators. Similarly no decrease is observed for stars with log($N_R$) $<$ $-$1.6.  Our results are in agreement with recent studies of the Ca IR triplet and thus no {\it super-saturation} exists in stellar chromospheres. However, as pointed out by many of these authors the true causes of saturation and super-saturation are still an open topic.

\begin{acknowledgments}
This research made use of the Cerro Tololo Observatory, 
National Optical Astronomy Observatory, which is operated by the
Association of Universities for Research in Astronomy, Inc. (AURA) 
under cooperative agreement with the National Science Foundation. 
The INT 2.5m is operated on the island of La Palma by the Isaac Newton Group in the Spanish Observatorio del Roque de los Muchachos of the Instituto de Astrof'sica de Canarias. This 
work was also supported by the UK Science and Technologies Facilities Council
(STFC).  We thanks Professor R. Jeffries for significantly improving the manuscript.
DC acknowledges support from the CSUN College of Science.
We would also like to thank I. Skillen  of ING NOAO for excellent support during our ING observations. 
This research has made use of the SIMBAD
database, operated at CDS, Strasbourg, France.
\end{acknowledgments}




%
%

 \begin{figure}
\epsscale{1.10}
\plottwo{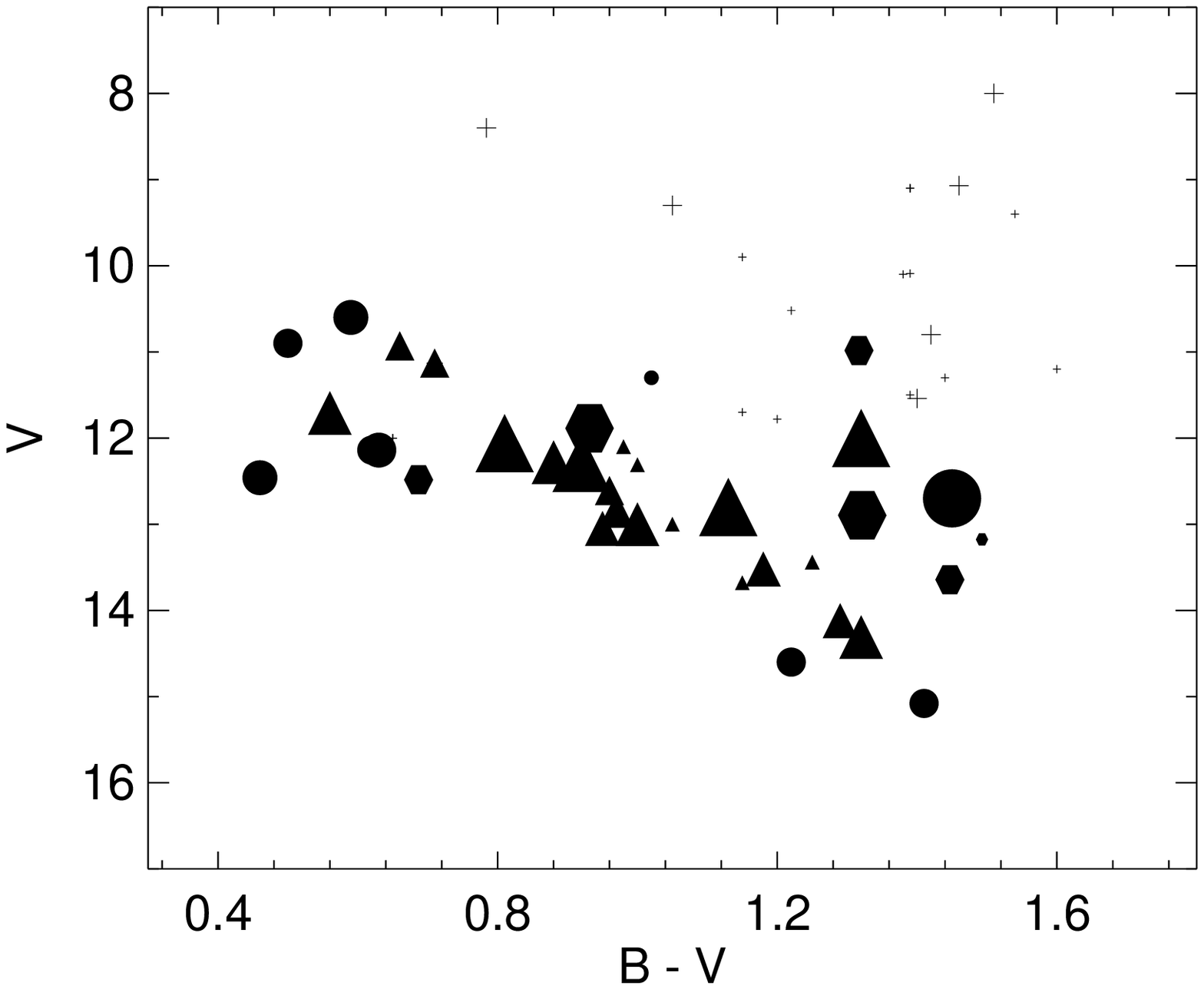}{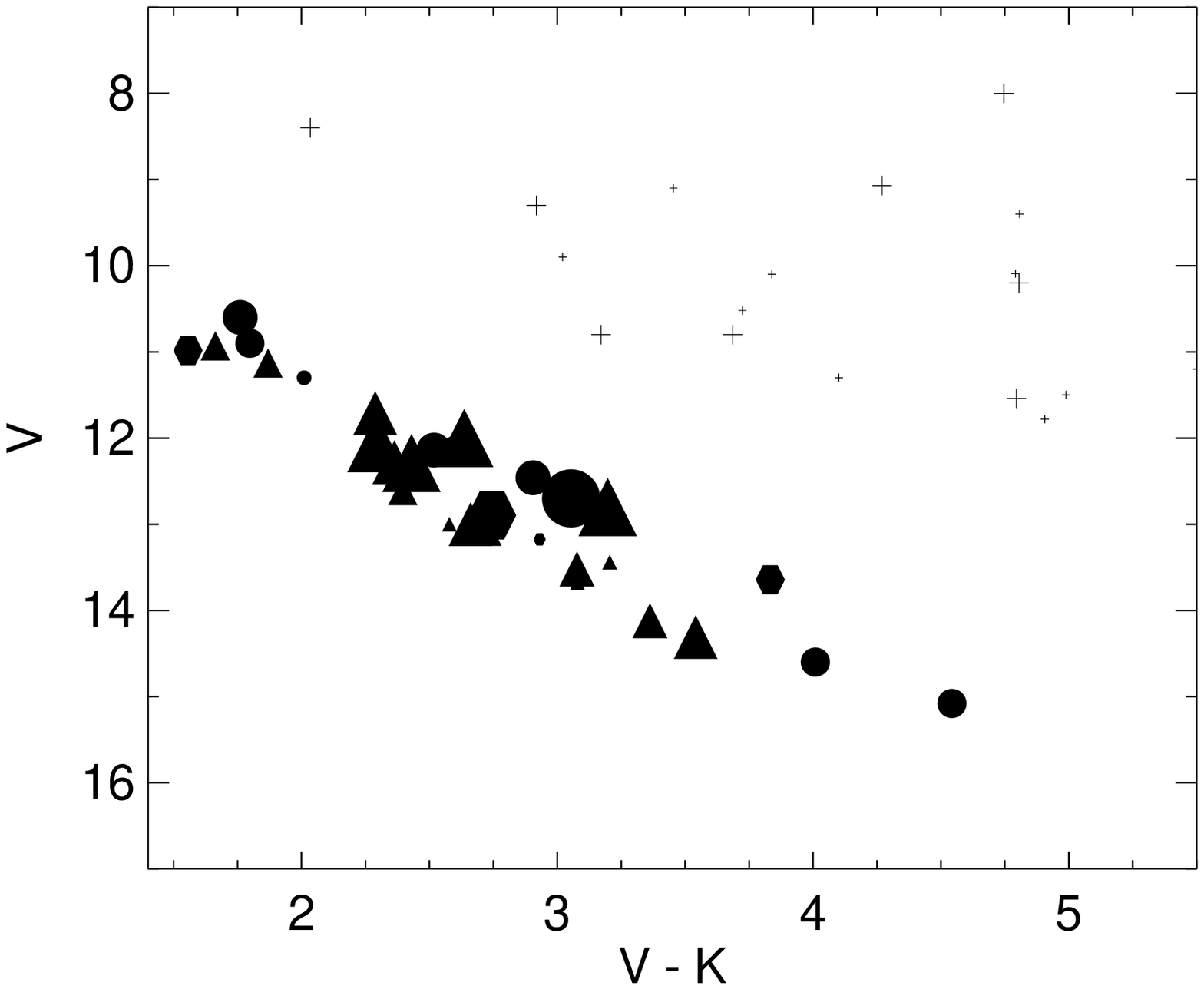}
\caption{Color-magnitude diagrams for our sample of stars. Shown in the left panel is V magnitude versus $B-V$, and V versus $V-K$ is shown in the right panel.  
Stars from $\alpha$ Per are shown as triangles,  the southern clusters, IC2391, IC2602 are shown as circles and hexagons, 
respectively, and miscellaneous {\it solar neighborhood} stars are shown as crosses.
Symbol size are proportional to rotational velocities, with the smallest symbols shown for vsini $<$ 25 km s$^{-1}$ and the largest for vsini $>$ 150 km s$^{-1}$. 
}
\label{fig-1}
\end{figure}

  \begin{figure}[t]
\epsscale{1.10}
\plottwo{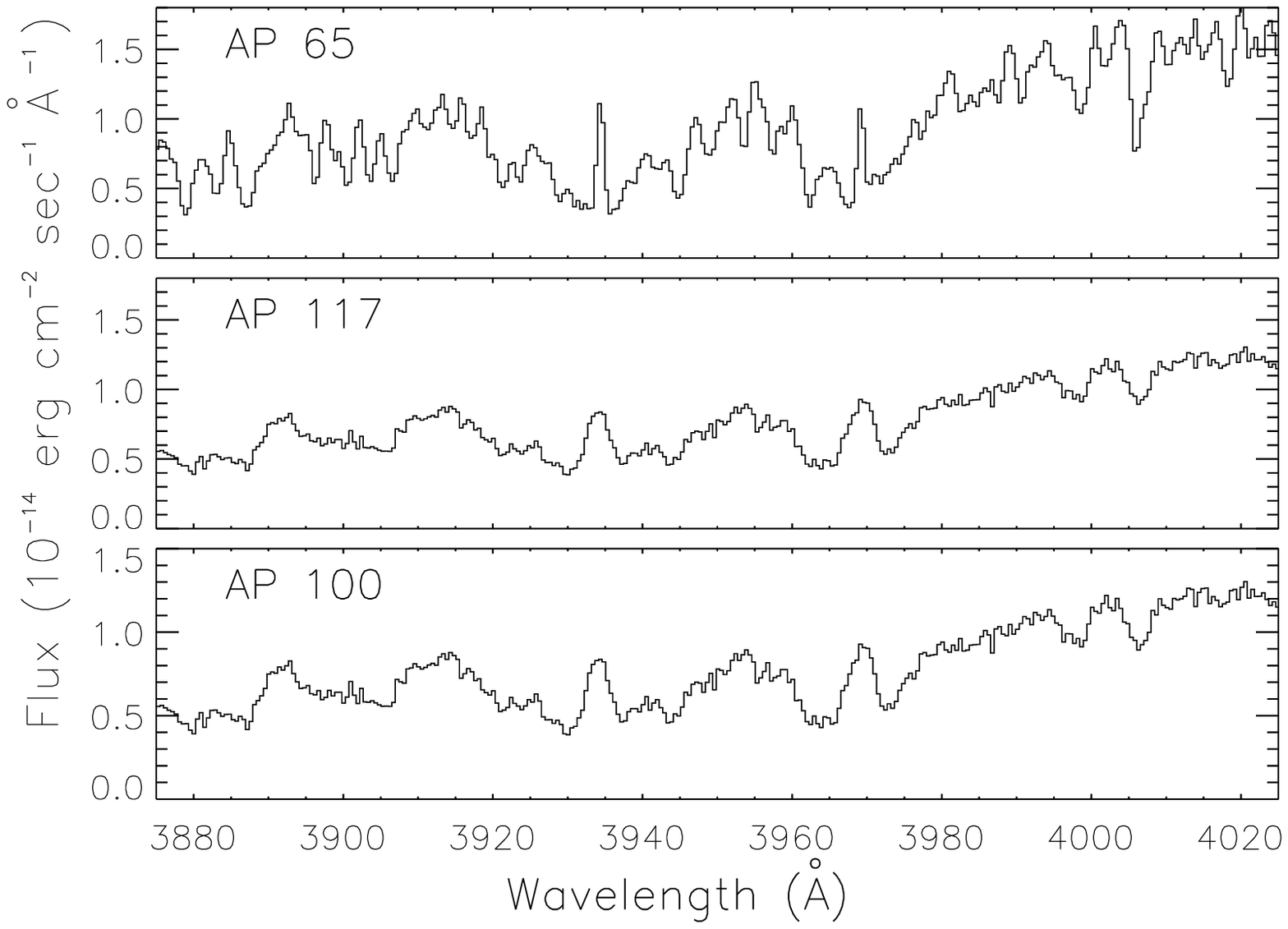}{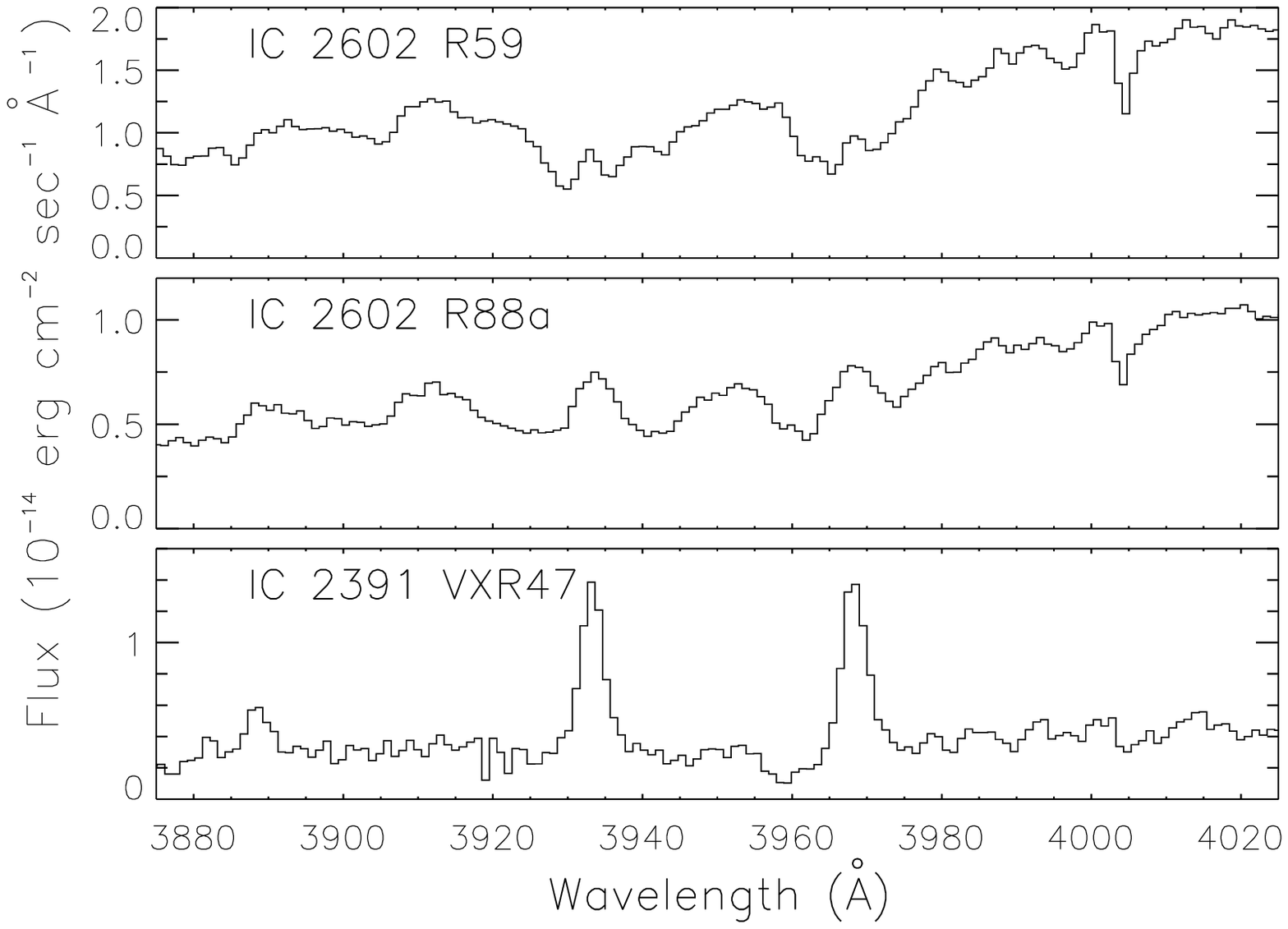}
\caption{Sample Ca II H \& K spectra.  The {\it left} panel shows sample INT IDS spectra for:
AP 65, a K star with rotational velocity of 10 km~s$^{-1}$, AP 117, 
a K star with a rotational velocity of 83 km~s$^{-1}$ and AP 100 an ultrafast rotating K7 star with vsini = 205 km~s$^{-1}$. Shown in the {\it right} panel are 
sample spectra taken with the R-C spectrograph at CTIO for: IC 2602 R59 a K0 star with a rotational velocity of 39 km~s$^{-1}$, IC 2602 R88a, an ultrafast rotating K4 star with $v$sini = 200 km~s$^{-1}$,
 and IC 2391 VXR 47, a fast rotating M star with $v$sini = 95 km~s$^{-1}$.
 }
\label{fig-2}
\end{figure}

\clearpage
\begin{figure}[t]
\epsscale{1.10}
\plottwo{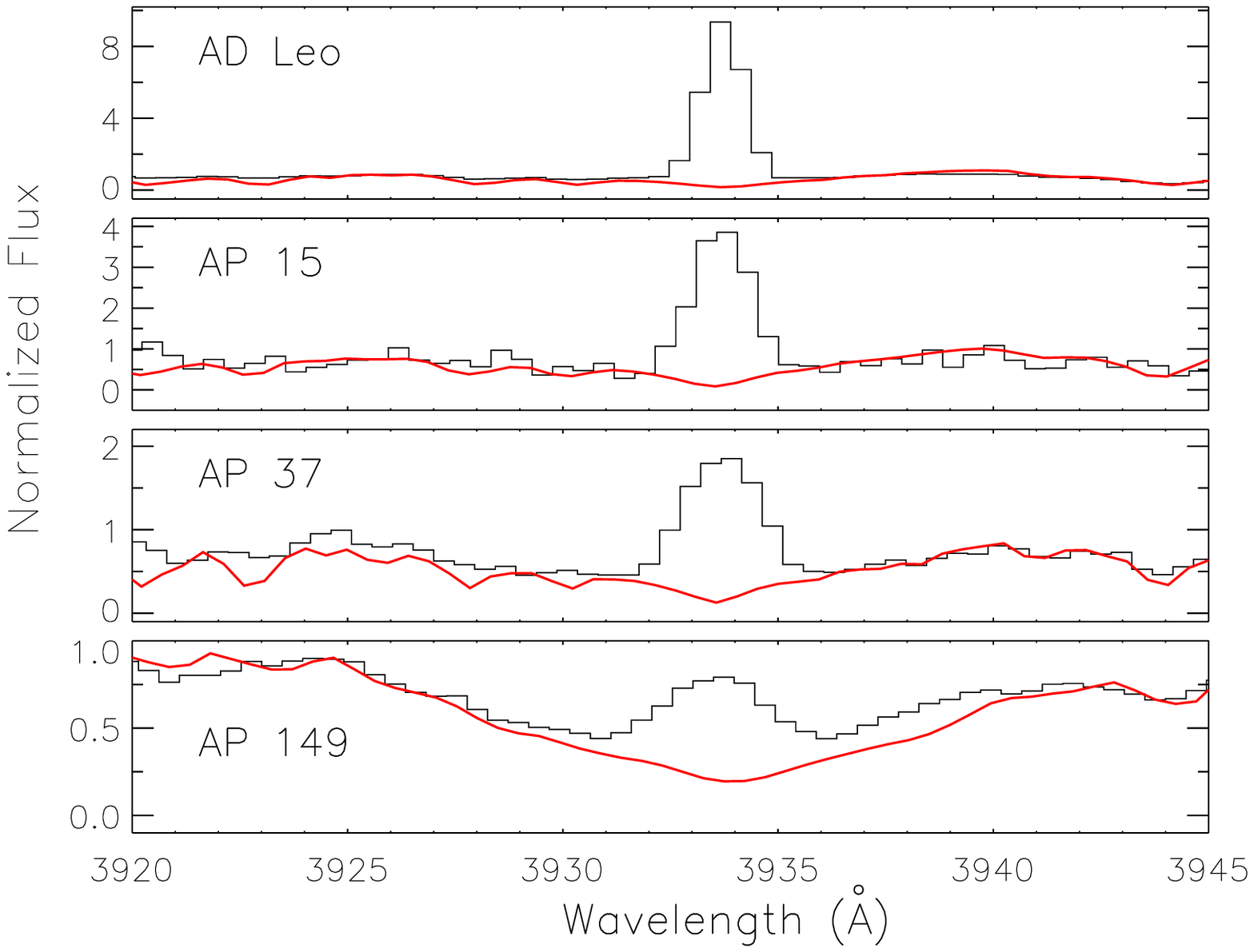}{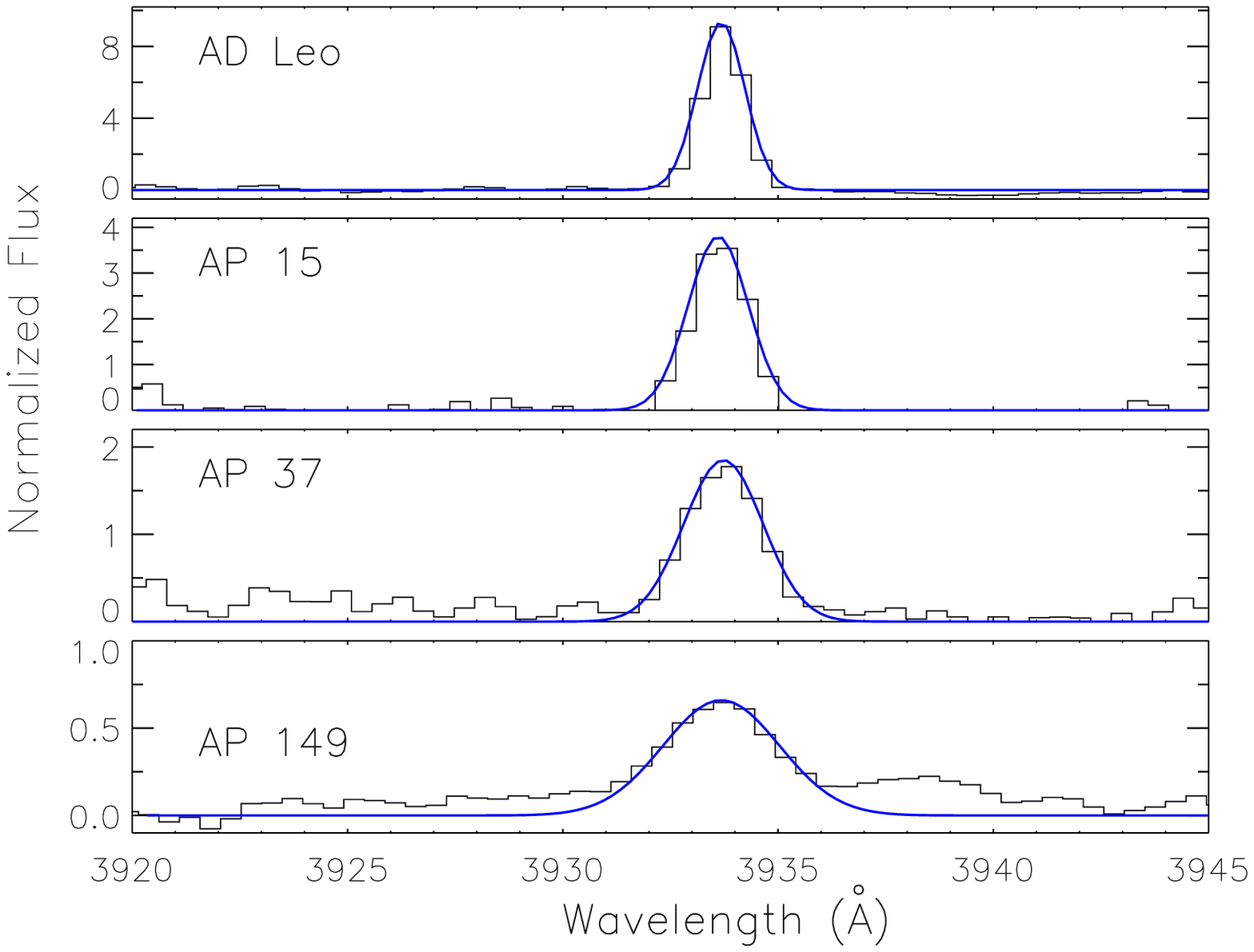}
\caption{Sample spectra including a synthesized photospheric model from SME. The {\it left} panel shows the observed spectra plotted as the black histogram and the SME model is over-plotted as the solid red line. The {\it right} panel shows the difference of the observed spectrum and SME model over-plotted with the fitted Gaussian as a solid blue line. 
Shown from top to bottom in both panels, are: AD Leo (T=3600K, vsini = 6 km~s$^{-1}$), AP 15 (T=4000, vsini = 52 km~s$^{-1}$), AP 37 (T=4900K, vsini = 29 km~s$^{-1}$), and
 AP 149 (T=6300,  vsini = 117 km~s$^{-1}$).
 }
\label{fig-3rr}
\end{figure}

\begin{figure}[t]
\epsscale{.90}
\plotone{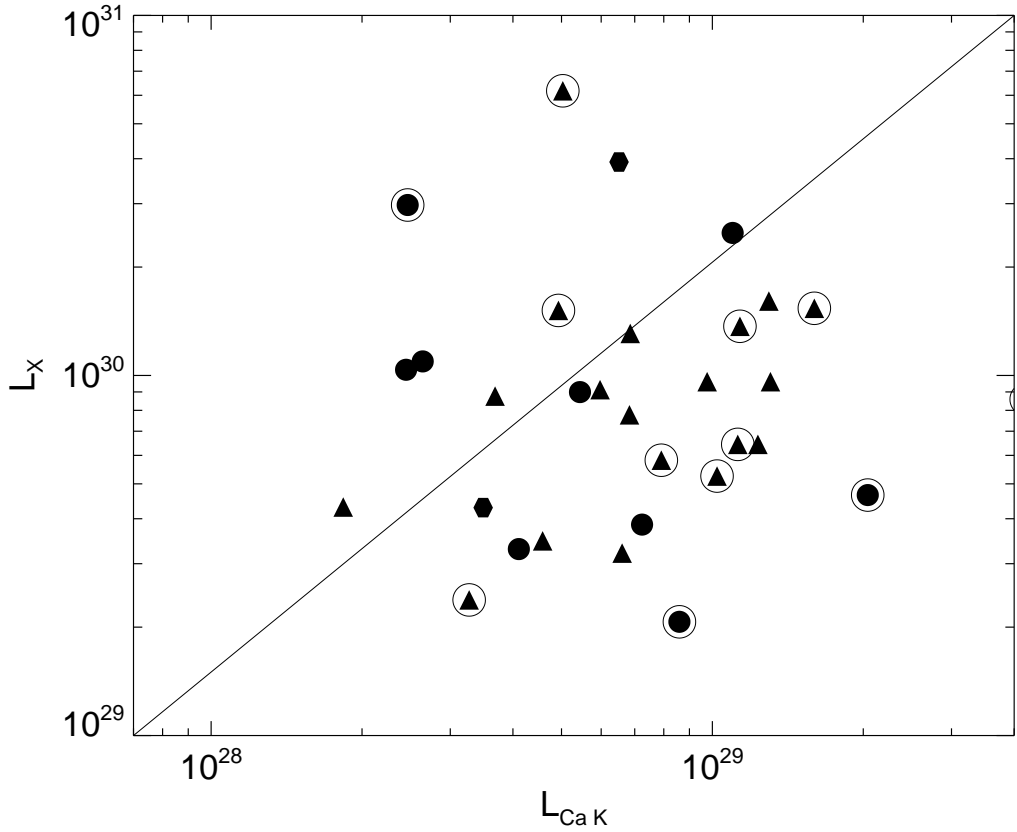}
\caption{The X-ray luminosity, $L_X$ plotted as a function of Ca K luminosities 
$L_{CaK}$ for cluster member stars. Stars showing super-saturation from James et al. (2000) are also indicated with an open circle. This 
shows a slight trend indicating that stars with supersaturation in the X-rays have higher  chromospheric emission (see text).  
Stars from $\alpha$ Per are shown as triangles,  the southern clusters, IC2391, IC2602 are shown as circles and hexagons, 
respectively.
}
\label{fig-4}
\end{figure}

\begin{figure}
\epsscale{1.00}
\plotone{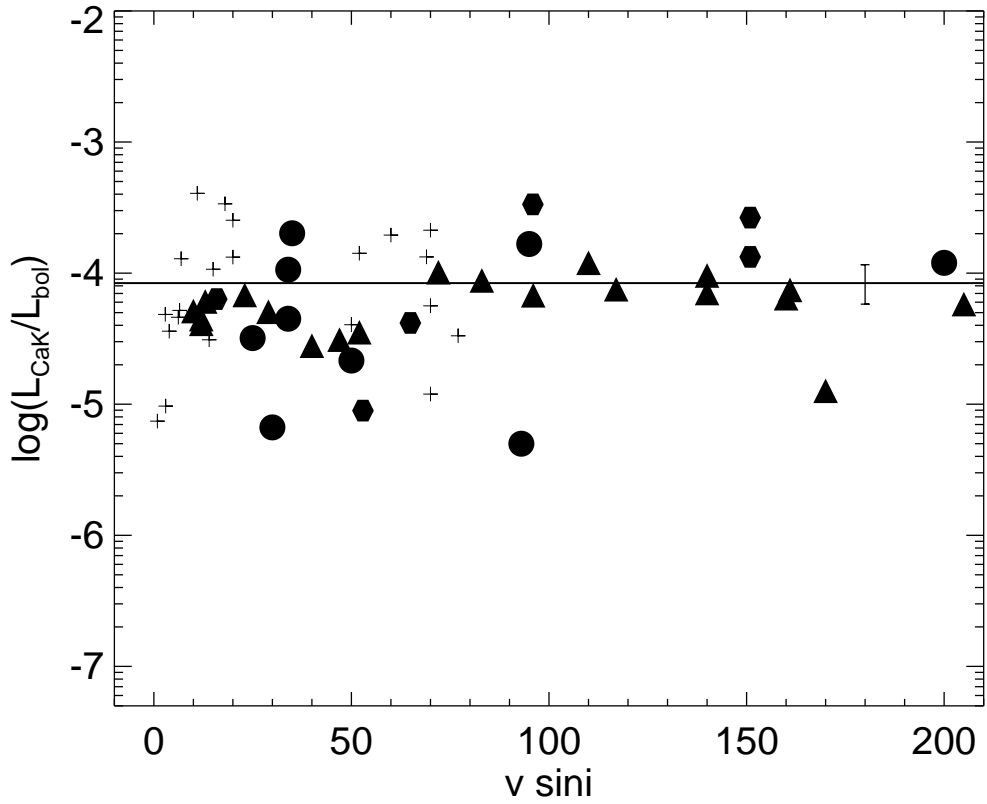}
\caption{Ca II K luminosities normalized to the bolometric luminosity as a function of rotational velocity. Symbols are the same as Figure 3,
 with the inclusion of the miscellaneous solar neighborhood stars shown as crosses.
 A line is shown for the mean $L_{CaK}/L_{bol}$ and a sample error bar is shown at v$sini$ = 180 km $s^{-1}$.
}
\label{fig-5}
\end{figure}

\begin{figure}
\epsscale{1.00}  
%
\plotone{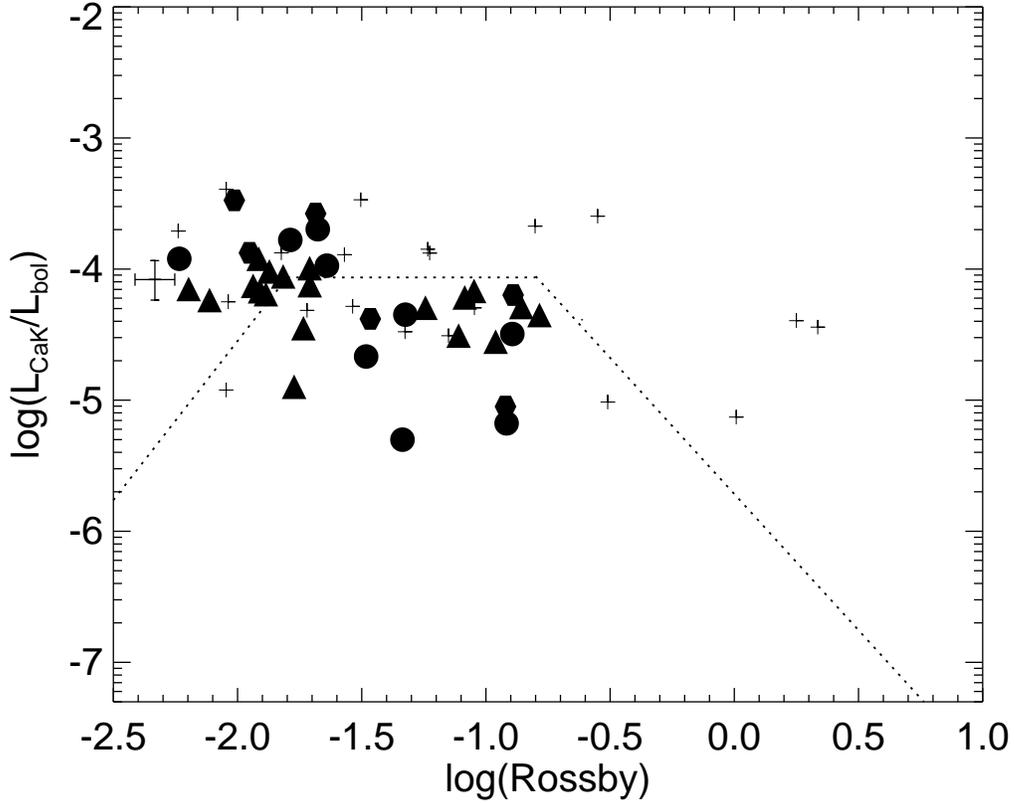}
\caption{Ca K luminosities normalized to the bolometric luminosity shown as a function of Rossby number, N$_R$ for all stars from our sample. 
Over-plotted are the luminosity levels and ranges from Jeffries et al. (2011) fitted to the median $L_{CaK}/L_{bol}$ showing saturation of
the X-rays and their decline at smallest Rossby numbers (see text). Symbols are the same as Figure 1, and a sample error bar is shown at log($N_R$) $\approx$ $-$2.3 for  $L_{CaK}/L_{bol}$ and log(N$_R$).
}
\label{fig-6}
\end{figure}









\begin{landscape}

\tabletypesize{\small}
\begin{deluxetable}{lccccccccccr}

\tablecolumns{14}
\tablewidth{0pc}
\tablenum{1}
\tablecaption{Ca H \& K Line Fluxes and Stellar Parameters} 
\tablehead{
\colhead{Source\tablenotemark{a}}  & \multicolumn{2}{c}{Line Fluxes\tablenotemark{b}}
   & \colhead{d} &\colhead{m$_V$}  &  \colhead{$B-V$} & \colhead{$V-K$} 
  & \colhead{$v$ sin$i$}  & \colhead{log(L$_{CaK}$/L$_{bol}$)\tablenotemark{b}} 
  & \colhead{log(L$_X$/L$_{bol}$)\tablenotemark{c}} & \colhead{$N_R$\tablenotemark{d}} & \colhead{Comment\tablenotemark{e}}\\
  \colhead{}  & \colhead{Ca K} & \colhead{Ca H} & \colhead{}
   & \colhead{} &\colhead{}  &  \colhead{} 
  & \colhead{} & \colhead{}  & \colhead{}  & \colhead{} & \colhead{}\\
\cline{2-3} \\
  \colhead{}
  & \multicolumn{2}{c}{(10$^{-14}$\,erg cm$^{-2}$ s$^{-1}$ \AA$^{-1}$ )}
 & \colhead{(pc)}  & \colhead{}   & \colhead{} & \colhead{} 
  & \colhead{(km s$^{-1}$)}   & \colhead{}  & \colhead{}   & \colhead{}}
\startdata

VXR60n&  5.2 &  5.9 &160.0 & 13.0 &  1.31 & 2.71 &  150& -3.90 & -3.52 &-1.97 & IC 2391  \\
VXR60s&  3.5 &  2.7 &160.0 & 13.2 & 1.49 & 2.92 &   15& -4.23 & -3.44 &-0.90 & S97 \\
VXR47&  5.0 &  4.9 &160.0 & 13.7 & 1.44 & 3.81 &   95& -3.50 &-3.36 &-2.03 & S97 \\
R50a&  1.1 &  0.5 &160.0 & 12.5  & 0.68 &-0.96 &   64& -4.41 & -3.32 &  -1.47& S97 \\
R62a&  2.1 &  2.2 &160.0 & 11.0 & 1.31 & 1.54 &   52& -5.08 &... &-0.94& S97 \\
R80a& 13.6 & 11.8 &160.0 & 12.0 &  0.92 &-1.79 &  150& -3.60 & -3.29 &  -1.70 & S97 \\
R3&  3.6 &  1.0 &160.0 & 11.3 & 1.02 & 2.01 &   25& -4.50 & -3.14 &-0.89 & IC 2602 \\
R24a&  1.3 &  2.1 &160.0 & 14.6 & 1.22 & 4.01 &   34& -3.97 & -3.07 &-1.64 & S97 \\
R31&  2.4 &  2.1 &160.0 & 15.1 & 1.41 & 4.54 &   35& -3.70 & -2.97 &-1.68 & S97 \\
R43&  0.9 &  0.4 &160.0 & 12.1 & 0.63 & 2.52 &   50& -4.67 & -3.05 &-1.48 & S97 \\
R52&  6.7 &  5.7 &160.0 & 12.5 & 0.46 & 2.90 &   95& -3.78 & -3.42 &-1.79 & S97 \\
R58&  0.8 & ... &160.0 & 10.6 & 0.59 & 1.76 &   93& -5.30 & -3.22 &-1.34 & S97 \\
R59&  1.8 &  1.5 &160.0 & 12.1 & 0.62 & 2.60 &   34& -4.35 & -3.13 &-1.32 & S97 \\
R83&  0.8 &  0.2 &160.0 & 10.9 & 0.50 & 1.80 &   30& -5.18 & -3.55 &-0.92 & S97 \\
R88a&  2.8 &  2.9 &160.0 & 12.7 & 1.45 & 3.05 &  200& -3.92 & -3.54 &-2.23 & S97 \\
J1131-34.6& 39.4 & 45.3 & 18.0 & 11.5 & 1.40 & 4.80 &   60& -3.71 & -3.34 &-2.24 & EUVE \\
J0825-16.3 & 44.8 & 53.4 & 46.0 & 10.5 & 1.22 & 3.72 &   20& -3.88 & -3.12 &-1.23 & EUVE \\
J1258-70.4& 11.7 &  9.0 &330.0 & 12.0 & 0.65 &-3.08 &   20& -3.60 & -3.40 &  -0.55 & EUVE \\ 
HD 174429&290.3 &162.8 &160.0 &  8.4 & 0.78 & 2.03 &   70& -3.67 &... &-0.80 & \\
GJ 431& 68.2 & 80.8 & 29.0 & 11.5 & 1.39 & 4.99 &   18& -3.47 & -3.01 &-1.50 & \\

AP 15&  0.5 &  0.5 &176.0 & 14.1 & 1.29 & 3.36 &   52& -4.45 & -3.08 &-1.73 & {\small $\alpha$ Per, R96}  \\  
AP 25&  1.9 &  1.3 &176.0 & 12.2 & 0.88 & 2.24 &   12& -4.35 & -3.07 &-0.78 & \\
AP 37&  1.6 &  1.0 &176.0 & 12.6 & 0.96 & 2.40 &   29& -4.29 & -3.11 &-1.24 &  {\small R96} \\ 
AP 43&  2.6 &  2.7 &176.0 & 12.8 & 0.97 & 2.73 &   72& -3.99 & -3.00 &-1.71 &  {\small R96} \\ 
AP 56&  2.8 &  2.6 &176.0 & 13.0 & 1.00 & 2.66 &  110& -3.92 & -3.21 &-1.91 & {\small R96} \\ 
AP 63&  3.0 &  2.7 &176.0 & 12.3 & 0.92 & 2.43 &  161& -4.13 & -3.37 &-1.94 &  {\small R96} \\ 
AP 65&  1.2 &  0.9 &176.0 & 13.0 & 1.05 & 2.58 &   10& -4.29 & -3.41 &-0.86 & {\small R96} \\ 
AP 86&  0.9 &  1.0 &176.0 & 14.3 & 1.32 & 3.54 &  140& -4.15 & -3.29 &-2.20& {\small R96} \\
AP 95&  3.9 &  2.4 &176.0 & 12.3 & 0.88 & 2.36 &  140& -4.02 & $<$ -3.29 &-1.87& {\small R96} \\
AP 100&  2.1 &  1.7 &176.0 & 12.8 & 1.13 & 3.20 &  205& -4.24 & -3.37 &-2.11& {\small R96} \\
AP 112&  1.0 &  0.9 &176.0 & 13.7 & 1.15 & 3.08 &   13& -4.22 & -2.84 &-1.09& {\small R96} \\
AP 124&  1.3 &  1.4 &176.0 & 13.4 & 1.25 & 3.20 &  ... & -4.18 & $<$ -3.36 &  ... & {\small R96} \\
AP 117&  1.8 &  1.8 &176.0 & 13.1 & 0.95 & 2.71 &   83& -4.06 & -3.00 &-1.82& {\small R96} \\
AP 118&  3.1 &  1.7 &176.0 & 12.1 & 0.81 & 2.29 &  160& -4.19 & -3.11 &-1.89& {\small R96} \\
AP 139&  1.3 &  0.5 &176.0 & 12.0 & 1.32 & 2.64 &  170& -4.90 & -3.41 &-1.77& {\small R96} \\
AP 149&  4.3 &  2.8 &176.0 & 11.7 & 0.56 & 2.29 &  117& -4.12 & -3.14 &-1.71 & {\small R96} \\
AP 167&  1.4 &  1.4 &176.0 & 13.5 & 1.18 & 3.08 &   96& -4.17 &-2.08 &-1.91 & {\small R96} \\
AP 197&  1.8 &  1.3 &176.0 & 12.3 & 1.00 &-3.01 &   12& -4.39 & -3.70 &  -0.62 & {\small R96} \\
AP 199&  3.5 &  2.4 &176.0 & 12.1 & 0.98 & 2.51 &   23& -4.17 &-3.30 &-1.05 & {\small R96} \\
HE 350&  3.3 &  2.1 &176.0 & 11.1 & 0.71 & 1.87 &   47& -4.51 & -3.79 &-1.11 & {\small R96} \\
HE 917&  3.5 &  2.3 &176.0 & 10.9 & 0.66 & 1.66 &   40& -4.55 & -3.46 &-0.96 & {\small R96} \\
HD 19305\tablenotemark{f} &  13.7 & 10.0 & 14.8 &  9.1 & 1.39 & 3.45 &   0& -5.13 &... & 0.01 \\
HD 17878&  0.7 &  1.0 & 74.0 &  3.9 & 0.73 &-4.49 &   25& -8.10 &... &  1.41 & \\
HD 283518& 10.7 & 11.1 &136.8 & 10.8 & -1.02 & 3.17 &   77& -4.48 &... & -1.33 &  \\
HD 283571& 39.9 & 18.4 &133.5 & 10.2 & 0.90 & 4.80 &   52& -3.95 &... & -1.23 &  \\
V1005 Ori& 22.1 & 20.9 & 26.7 & 10.1 & 1.38 & 3.84 &   14& -4.51 &... &-1.15 & \\
J0723+20& 26.9 & 33.2 & 25.8 &  9.9 & 1.15 & 3.02 &   10& -4.30 &... &-1.05 & RE \\
YZ CMi& 25.9 & 30.5 &  5.9 & 11.2 & 1.60 & 5.50 &    6& -4.28 &... &-1.54 & \\
REJ0808+21& 17.0 & 11.3 &  6.0 & 11.7 & 1.15 & 5.62 &  ...& -3.78 &... &  ... & RE \\
CV Cnc&  1.9 &  1.9 & 12.8 & 13.6 & 1.50 & 5.88 &    6& ... &... & ... & \\
J1004+50& 31.0 & 34.3 & 14.0 & 11.3 & 1.44 & 4.10 &   15& -3.97 &... &-1.67 & EUVE \\
GJ 380&173.6 &130.2 &  4.9 &  6.6 & 1.38 & 3.64 &    3& -5.01 &... &-0.51 & \\
AD Leo&103.1 & 92.7 &  4.7 &  9.4 & 1.54 & 4.81 &    6& -4.34 & -3.21 &-1.31 & \\
CN Leo&  5.0 &  4.8 &  2.3 & 13.5 & 2.01 & 7.42 &    2& -4.32 &... &-1.72 & \\
GJ 411 & 302.0 & 136.0&     2.5 &   8.0 &  1.51&  4.75 &    50 & -4.39 & -4.30  & 0.25 & \\  
HD 131156& 1624.0 &  950.0 &   25.0 &    4.6 & 0.77 & 2.63 &    3& -4.44 &... & 0.34 & \\
J2131+23& 96.7 & 93.9 & 25.1 &  9.3 & 1.05 & 2.92 &   69& -3.93 &... &-1.82 & EUVE \\
EV Lac& 95.3 &127.3 &  5.1 & 10.1 & 1.39 & 4.79 &    6& -3.89 &... &-1.57 & \\
FK Aqr& 35.2 & 35.9 &  8.6 &  9.1 & 1.46 & 4.27 &   70& -4.92 &... &-2.05 & \\
GT Peg& 10.5 & 10.4 & 14.3 & 11.8 & 1.20 & 4.91 &   11& -3.39 &... &-2.05 & \\
GJ 890& 24.2 & 31.1 & 22.0 & 10.8 & 1.42 & 3.69 &   70& -4.25 & -3.18 &-2.04 & \\
      \enddata
        \tablenotetext{a}{Star names using the convention of Stauffer et al. 1997 for IC2391 and IC 2602, and Randich  et al. 1996 for $\alpha$ Per. }
\tablenotetext{b}{Ca H \& Ca K Fluxes and L$_{CaK}$/L$_{bol}$ from the current work.}
\tablenotetext{c}{X-ray luminoisty, L$_{X}$/L$_{bol}$ and vsini references with: S97 = Stauffer et al. 1997 for IC 2391 and IC 2602,\\ unless otherwise noted. R96 = Randich et al. 1996 for $\alpha$ Per (AP); 
J00 = James et al. 2000. }
\tablenotetext{d}{N$_R$ Rossby number \citep{J11}}
\tablenotetext{e}{Comment of Star Cluster Name or abbreviated star name; RE for ROSAT EUV and EUVE\\ for Extreme Ultraviolet Explorer from Christian \& Mathioudakis 2002.} 
\tablenotetext{f}{Start of solar neighborhood stars}

\end{deluxetable}
\end{landscape}


\end{document}